# Ultrafast Graphene Light Emitter

Young Duck Kim[1]†*, Yuanda Gao[2]†, Ren-Jye Shiue[3], Lei Wang[4], Ozgur Burak Aslan[5], Myung-Ho Bae[6,7], Hyungsik Kim[8], Dongjea Seo[9], Heon-Jin Choi[9], Suk Hyun Kim[5], Andrei Nemilentsau[10], Tony Low[10], Cheng Tan[2], Dmitri K. Efetov[3,11], Takashi Taniguchi[12], Kenji Watanabe[12], Kenneth L. Shepard[8], Tony F. Heinz[5], Dirk Englund[3], and James Hone[2]*

[1]Department of Physics, Kyung Hee University, Seoul 02447, Republic of Korea.

[2]Department of Mechanical Engineering, Columbia University, New York, NY 10027, USA.

[3]Deparatment of Electrical Engineering and Computer Science, Massachusetts Institute of Technology, Cambridge, MA 02139, USA.

[4]Kavli Institute at Cornell for Nanoscale Science, Ithaca, NY 14853, USA.

[5]Department of Applied Physics, Stanford University, Stanford, CA 94305, and SLAC National Accelerator Laboratory, Menlo Park, CA 94025, USA.

[6]Korea Research Institute of Standards and Science, Daejeon 34113, Republic of Korea.

[7]Department of Nano Science, University of Science and Technology, Daejeon 34113, Republic of Korea.

[8]Department of Electrical Engineering, Columbia University, New York, NY 10027, USA.

[9]Department of Materials Science and Engineering, Yonsei University, Seoul 120-749, Republic of Kore.

[10]Department of Electrical and Computer Engineering, University of Minnesota, Minneapolis, MN 55455, USA.

[11]ICFO-Institut de Ciencies Fotoniques, The Barcelona Institute of Science and Technology, 08860 Castelldefels, Barcelona, Spain.

[12]National Institute for Materials Science, 1-1 Namiki, Tsukuba, 305-0044, Japan.

†These authors contributed equally to this work.

*Correspondence to: ydk@khu.ac.kr, jh2228@columbia.edu

**Abstract**: Ultrafast electrically driven nanoscale light sources are critical components in nanophotonics. Compound semiconductor-based light sources for the nanophotonic platforms have been extensively investigated over the past decades. However, monolithic ultrafast light sources with a small footprint remain a challenge. Here, we demonstrate electrically driven ultrafast graphene light emitters that achieve light pulse generation with up to 10 GHz bandwidth, across a broad spectral range from the visible to the near-infrared. The fast response results from ultrafast charge carrier dynamics in graphene, and weak electron-acoustic phonon-mediated coupling between the electronic and lattice degrees of freedom. We also find that encapsulating graphene with hexagonal boron nitride (hBN) layers strongly modifies the emission spectrum by changing the local optical density of states, thus providing up to 460 % enhancement compared to the grey-body thermal radiation for a broad peak centered at 720 nm. Furthermore, the hBN encapsulation layers permit stable and bright visible thermal radiation with electronic temperatures up to 2,000 K under ambient conditions, as well as efficient ultrafast electronic cooling *via* near-field coupling to hybrid polaritonic modes. These high-speed graphene light emitters provide a promising path for on-chip light sources for optical communications and other optoelectronic applications.

Intense research over the past decades has focused on the development of high bandwidth photonics for inter-/intra-chip connections and other applications, with a specific aim at nanophotonic building blocks such as waveguides, optical modulators, and photodetectors. However, on-chip light sources, particularly monolithic nanoscale light sources with direct high-speed modulation, have remained challenging [1]. Due to its unique electronic and optical properties, graphene has emerged as a promising material for optoelectronic applications, including as ultrafast and broadband photodetectors [2,3], optical modulators [4,5], plasmonics [6-8] and nonlinear photonic devices [9]. Previous graphene devices have shown the feasibility of ultrafast signal processing and frequency conversion functionalities required for photonic integrated circuits [9,10].

Graphene's high thermal stability, low heat capacity, and ultrafast opto-electronic properties [3,11] suggest that it could function as an unusual fast and efficient thermal light emitter. Early efforts showed infrared light emission from $SiO_2$-supported graphene, with temperatures limited to ~ 1,100 K [12-14] due to dielectric degradation at high temperature [15] and significant hot carrier cooling to the substrate. We recently demonstrated thermal light emission in the visible range from electrically biased suspended graphene [16] which achieves temperature up to ~ 2,800 K. However, to achieve rapid cooling required for fast modulation and to integrate such devices into photonic platforms, a substrate-supported device design is needed. Moreover, little is known about the possible modulation rate of graphene thermal emitters.

Here, we demonstrate electrically driven ultrafast thermal light emitters based on hBN-encapsulated graphene. The hBN allows roughly 60% larger current density than for $SiO_2$-supported graphene due to its larger optical phonon energy [17] and, at the same time, provides excellent encapsulation. As a result, our devices achieve electron temperatures up to 2,000 K and produce broadband emission extending up to the visible range. Our studies further indicate device lifetimes of years in vacuum and good stability even under ambient conditions. The thermal emission spectrum is strongly modified by the hBN dielectric optical cavity [5,18], which provides 460% enhancement for a broad peak centered at 720 nm compared to the grey-body thermal radiation. Analysis of thermal transport in the devices shows that the hBN effectively spreads heat over the micron scale, and that the dominant thermal transport pathway is vertical, with good agreement between models and the measured power consumption and temperature profile. Independent measurements of electron and acoustic phonon temperatures indicate that the electrons can be ~30% hotter than the acoustic phonons at high bias due to weak electron-acoustic phonon coupling [11,19]. Studies of the light emission under radiofrequency and pulsed excitation show continuous modulation at 3 GHz and emission pulses of 92 ps full width at half maximum (FWHM). This observation is consistent with a model in which electrons are strongly coupled to hybrid plasmon-phonon polaritonic modes at the graphene-hBN interface, but out of equilibrium with the acoustic phonons.

To fabricate the graphene light emitters, hBN/graphene/hBN heterostructures were first assembled by a van der Waals dry pick-up method using exfoliated monolayer graphene and exfoliated hBN flakes with 10-20 nm thickness and transferred to a $SiO_2$ (285 nm)/Si substrate, as shown in Fig. 1A. Electrical contacts were formed by etching the assembled heterostructure and depositing metal (Cr/Pd/Au) on the exposed edge [20]. The resulting graphene heterostructure exhibits mobility near the intrinsic acoustic phonon scattering limit at room temperature [20]. The atomically clean interface reduces extrinsic effects [21,22] such as surface roughness, defects and charged impurities. This permits investigation of intrinsic electro-thermal properties, including thermal radiation, energy dissipation, and ultrafast dynamics of hot electrons in the disorder-free graphene system.

Under high electric fields ($F$) up to ~ 6.6 V/μm and zero back gate voltage ($V_{BG}$), these devices achieve current density ($J$) up to ~ $4.0 \times 10^8$ A/cm$^2$. This high current density is due both to the high stability of the hBN and its high optical phonon energy, as will be explored further below. At high current density, we observe remarkably bright visible light emission from these micron-scale structures - even observable by the naked eye, as shown in Fig. 1C. The emission is seen across the channel region and increases in intensity with $F$ as shown in Figs. 1D and 1E (see Supporting Information Movie Clips).

Because stability is essential for practical applications, we tested the long-term performance of the graphene light emitter under high electric field ($F$ = 4.2 V/μm) and high current density ($J \sim 3.4 \times 10^8$ A/cm$^2$) under ~ $10^{-5}$ Torr vacuum. These measurements showed no significant degradation of emission intensity and electrical current over a test period of ~ $10^6$ seconds as shown in Fig. 1F, suggesting a device lifetime (defined by 50% degradation in current) exceeding 4 years. This result attests to the remarkable stability of both the hBN encapsulation [23,24] and edge contacts even under high electric field, current density, and temperature. Important for practical applications, we also observed visible light emission under ambient conditions: the best devices showed stable operation in air for several days, and it is likely that improved encapsulation will extend this lifetime.

Figure 2A shows the spectrum of the emitted light for a range of applied electric fields (or electric powers) under vacuum conditions. The spectrum extends from the visible to near-infrared (400 ~ 1,600 nm), with an emission peak around 720 nm and a flat response in the near-infrared (> 1,000 nm) from several graphene light emitters. The spectrum is unchanged for emission in air (Fig. 2B). The strong emission peak at 720 nm from the hBN encapsulated graphene light emitter can be attributed to the formation of a dielectric optical cavity by the hBN layers (refractive index $n = 2.2$), and the resulting tailoring of thermal radiation by the modified local optical density of states [25]. The intensity of thermal radiation from graphene at a given angle $\theta$, defined with respect to the normal to graphene surface, was calculated using the generalized Kirchoff's law [26],

$$I_{\omega,\alpha}(\omega, \theta, T_e) = a_{\omega,\alpha}(\omega, \theta, T_e) I_{\omega,b}(\omega, T_e),$$

where $a_{\omega,\alpha}(\omega, \theta, T_e)$ is a spectral directional absorptivity (emissivity) of the graphene layer in the stack for a given polarization of electromagnetic wave $\alpha = TE, TM$, $\omega$ is frequency, $I_{\omega,b}(\omega, T_e) = \omega^2 \Theta(\omega, T_e)/8\pi^3 c^2$ is the intensity of blackbody radiation for a single polarization, $\Theta(\omega) = \hbar\omega/(\exp(\hbar\omega/k_B T_e) - 1)$, $\hbar$ is the reduced Planck's constant, $k_B$ is the Boltzmann's constant, and $T_e$ is the electron temperature, which is used as a fitting parameter. The absorptivity (emissivity) of the graphene was calculated analytically by solving Maxwell's equations for a plane wave incident on the hBN/graphene/hBN on SiO$_2$/Si substrate. This analysis reproduces the data well (solid lines in Fig. 2B) and implies $T_e = 1,980$ K for $F = 5.0$ V/μm. The radiation enhancement due to the hBN layers reaches 460 % at the 720 nm peak, relative to graphene grey-body thermal radiation at same $T_e$ [25] as detailed in Supporting Information. We note that a flat response in the near-infrared (> 1,000 nm) is attributed to the Pauli blocking reduce absorptivity (emissivity) of graphene at high $T_e$ with increased Fermi energy.

Figure 2C shows that the derived $T_e$ increases roughly linearly with applied electrical power density ($P_e$), indicating that the dominant heat transfer mechanism is by conducting through the substrate rather than radiation. Consistent with this observation, dividing the total output optical power ($P_r$) across all wavelength based on the Stefan-Boltzmann law by $P_e$, we find an radiation efficiency $\eta \sim 3.45\times10^{-6}$ (see Supporting Information). This is smaller than obtained for suspended graphene, but can be improved by optical and thermal engineering.

To gain insight into the thermal transport processes that determine the efficiency, lateral extent, and dynamics of the light emission, we measured both the electronic and vibrational (acoustic phonon) temperature of the light emitters. To provide a second measurement of $T_e$ in addition to that obtained by fitting the radiation spectra (Fig. 2B), we analyzed the high-bias electrical transport behavior. The *I-V* behavior for different values of the $V_{BG}$ (Fig. 3A) showed current saturation [16,17,27] under modest electrical fields ($F > 0.5$ V/µm) for $V_{BG} > 20$ V. This saturation can be attributed to efficient backscattering of electrons by emission of strongly coupled optical phonons when the optical phonon activation length $L_\Omega (\propto \hbar\Omega/F$, where $\hbar\Omega$ is the optical phonon energy) [28] becomes smaller than the acoustic phonon scattering length $L_{ap}$, which approaches 1 µm in a hBN encapsulated graphene at room temperature [20]. In this regime, hot electrons in graphene can emit optical phonons in both the graphene and the hBN ($\hbar\Omega \sim 150\text{-}200$ meV). $L_\Omega$ approaches 500 nm at $F > 0.3\text{-}0.4$ V/µm, consistent with the observed onset of current saturation. In SiO$_2$–supported devices, hot electrons can emit SiO$_2$ optical phonons with lower energy ($\hbar\Omega_{SiO_2} \sim 60\text{-}80$ meV), resulting in a lower current density (Figure S4). This highlights the importance of the hBN for achieving efficient visible light emission. Moreover, consistent with optical studies which show electron-optical phonon scattering on a sub-100 fs time scale[3], this result indicates that electrons are strongly coupled with optical phonon modes of the graphene and hBN.

We measured $T_e$ by plotting sheet conductance ($\sigma$) against $V_{BG}$ for different values of $F$ (Fig. 3B). For $F < 3.3$ V/µm, $\sigma$ is clearly modulated by $V_{BG}$, while above 4 V/µm, $\sigma$ is nearly independent of $V_{BG}$. This reduction in gate modulation is due to the interplay between electrostatically induced charge carriers $n_g \propto V_{BG}$ and thermally generated charge carriers, $n_{th} \propto T_e^2$ [16,29,30]. Since $\sigma \propto n_{tot}e\mu$, where $n_{tot} = n_{th} + n_g$ and $e$ is the electron charge, the gate modulation effect becomes small when $n_{th} \gg n_g$. Using $\mu \propto T_e^{-\beta}$ for the temperature-dependent mobility, where $\beta = 2.5$ is obtained from a numerical self-consistent heat transport model, and using $T_e$ as an adjustable parameter, we performed numerical calculations of the graphene self-heating that show good agreement with the measured data (Fig. 3B, solid lines) (see Supporting Information for details). The derived values of $T_e$ are close to those obtained from fitting the radiation spectrum (Fig. 2B).

We next measured $T_{ap}$ of graphene and hBN by Raman spectroscopy: the graphene G mode and the hBN $E_{2g}$ mode shift downward with increasing $T_{ap}$ due to anharmonic phonon coupling [31] (see Supporting Information for detail). Figure 3C shows the variation of these modes up to $F = 3.7$ V/μm, above which the visible radiation background interfered with the measurement. The derived temperatures, together with the electronic temperature derived above, are shown in Fig. 3D. $T_{ap}$ of the graphene and hBN are nearly equal, but fall below $T_e$ at high bias. Thus, these measurements indicate that $T_e$ is out of equilibrium with $T_{ap}$ due to the energy relaxation bottleneck, which has been seen to follow $T_e = T_{ap} + \alpha(T_{ap} - T_0)$, where $\alpha$ is a non-equilibrium temperature coefficient and $T_0$ is the ambient temperature [16,29,30]. Based on the measured $T_e$ and $T_{ap}$ in the hBN encapsulated graphene heterostructure, we find $\alpha \sim 0.45$-$0.77$. This result is consistent with ultrafast spectroscopic studies which have revealed that hot electrons thermalize rapidly through strong electron-electron and electron-optical phonon scattering [3,11,19], but more slowly with acoustic phonons [11,19,29]. In addition, previous measurements have indicated that electron and acoustic phonon temperatures can be out of equilibrium under high electrical bias [16,19]. However, given the uncertainty in calibration of the Raman shift rates with temperature and possible confounding effects such as substrate thermal expansion, this result alone is not sufficient to definitively establish the disequilibrium between $T_e$ and $T_{ap}$.

Because heat dissipation occurs primarily through transport of acoustic phonons, we used the measured lattice temperature ($T_{ap}$ ~ 1,250-1,450 K, which corresponds to $T_e$ ~ 2,000 K) to calculate the total thermal resistance $R_{th}$ to the substrate by $T_{ap} - T_0 = R_{th} P_e$ [29]. This calculation yields $R_{th}$ ~ 10,650-11,480 K/W, which matches reasonably well with a simple model in which heat flow is dominated by vertical transport through the hBN and SiO$_2$ to the Si substrate [32] (See Supporting Information and Table S1). This analysis show that $R_{th}$ is dominated by thermal resistance of SiO$_2$ layer (~ 8,000 K/W, for 285 nm thickness).

We also calculated the $T_e$ distribution in the hBN encapsulated graphene light emitter based on the non-equilibrium temperature coefficient $\alpha$ and heat diffusion equation of $T_{ap}$ (see Supporting Information). Combining the vertical thermal transport results above with the lateral thermal conductivity of the hBN allows calculation of the lateral thermal diffusion (healing) length ($L_H$ ~ 1.3 µm), which is similar to the observed size of the bright emission seen in Fig. 1E. Fig. 3E plots the resulting $T_e$ distribution along the graphene light emitter for various values of $F$. In all cases, the cooling to the substrate keeps $T_e$ near the metal electrodes below ~ 600 K, explaining the high stability of the devices. The expected thermal radiation intensity profile based on the modeled temperature distribution is shown in Fig. 3F, in good agreement with the measured optical intensity profile (Fig. 3F inset, and Supporting Information). This analysis suggests how to optimize the device design by thermal engineering of the vertical thermal conductivity and lateral device dimensions.

The small size and low heat capacity of the graphene emitter presents an opportunity for ultrafast thermal emission modulation. Moreover, measurement of the dynamics of light modulation under fast pulses may provide insight into the carrier dynamics and offer another means to examine whether electron and phonon populations are out of equilibrium under high bias. Moreover, recent studies [33,34] also suggest that direct electronic cooling into hBN can be mediated by efficient near-field heat transfer due to the hybrid plasmon-phonon polaritonic mode at highly localized graphene/hBN interface [34]. Therefore, we examined the ultrafast response of a device fabricated on a quartz substrate, which reduces parasitic capacitance and enables electrical driving at GHz frequencies with DC offset bias ($V_{DC}$) (Fig. 4) (the device exhibits identical steady state radiation as observed above for the $SiO_2$/Si substrate-mounted devices). As a first test, the emission time trace in Fig. 4B shows on-off modulation with near-perfect contrast when device is driven with a continuous 3 GHz signal. For an even shorter pulse duration (FWHM 80 ps, peak to peak 2V with $V_{DC}$ =1.6V), the output light pulse width is only broadened to 92 ps (FWHM), which corresponding to above 10 GHz bandwidth as shown in Fig. 4C. This response is many orders of magnitude faster than conventional thermal radiation sources based on bulk materials, for which modulation speed has been limited to ~ 100 Hz [35].

This observed ultrafast response may arise from the small size and thermal mass of the graphene, since vertical thermal diffusion can occur over sub-ns timescales for nm-scale structures. More intriguingly, if electrons are out of equilibrium with the acoustic phonons as indicated in Fig. 3, this high speed may be due to ultrafast cooling from $T_e$ to $T_{ap}$, which should be sufficient to modulate the output intensity by direct electronic cooling mediated with near-field heat transfer *via* hybrid plasmon-phonon polaritonic mode at graphene/hBN interfaces.

The simple heat transfer model shown in the inset of Fig. 4C allows us to quantify the transient cooling behavior. In this model, we assume that the hot graphene electrons are strongly coupled to and in equilibrium with the optical phonons of graphene and the top few layers of the hBN [19,34]. These optical phonons are connected to the acoustic phonon bath by thermal conductance $\Gamma_E$ and to the environment by $\Gamma_0$ [36]. $\Gamma_E$ can be derived from non-equilibrium temperature ($T_e \sim 2,000$ K, $T_{ap} \sim 1,000$-$1,280$ K) under steady state measurements shown in Fig. 3D: $T_e - T_{ap} \approx P/\Gamma_E$ [36], with $\Gamma_E \sim 6.0 - 8.4$ MWm$^{-2}$K$^{-1}$, which is consistent with theory [37]. The fast cooling time constants ($\tau_c$) of graphene light emitter was estimated by fitting based on the $I \propto \exp[-\hbar\omega_{peak}/k_B(T_0 + dT\exp(-t/\tau_c))]$, where $\hbar\omega_{peak}$ is the maximum spectral sensitivity of avalanche photodetector (corresponding to the 550 nm), $T_0$ is offset temperature controlled by DC offset bias, and $dT$ is modulated temperature by RF pulses. Based on the fitting of time-resolved light pulse output from graphene with $T_0$ (300-700 K) and $dT$ (20-200 k) parameters, we obtained $\tau_c \sim$ 120-900 ps (see Supporting Information). Estimated $\tau_c$ allow us to determine the heat capacity of the electron/optical phonon system, $C_T = \tau_c \Gamma_E = 0.72 - 6.63 \times 10^{-3}$ Jm$^{-2}$K$^{-1}$. This value is $10^4$ times larger than the electronic heat capacity of graphene (see Supporting Information), confirming the assumption that optical phonons are in equilibrium with the electrons. We find that the magnitude of $C_T$ corresponds to the optical phonons in graphene in addition to 0.3-3.6 nm of the surrounding hBN [38,39]. This view is consistent with theoretical predictions for hybrid modes that are highly localized at the graphene-hBN interface [33,34].

This work establishes that hBN-encapsulated graphene provides visible light emission with high stability and a modulation rate speed several orders of magnitude faster than conventional thermal emitters. Our quantitative model predicts that the exceptional speed likely arises because hot electrons are strongly coupled to optical phonons and hybrid plasmon-hyperbolic phonon polariton modes in hBN, but weakly coupled to acoustic phonons, resulting in disequilibrium between the two populations, though the dynamics of this process need to be studied in more detail. We found that the emission spectrum is strongly modified by a tailored density of optical modes in the hBN slab; this modification may be engineered to sharply reshape the emission spectrum by coupling the hBN encapsulated graphene to an optical cavity. Furthermore, making use of a tunable energy relaxation pathway for the graphene light emitter, such as tunneling structures [40], could allow an even faster light modulation rate beyond the speed limits explored here. Finally, we note that graphene thermal emitters employ the same basic device architecture as demonstrated previously for ultrafast photodetectors and electro-optic modulators [2,5,8]. Thus, one graphene-hBN heterostructure device could serve three essential electro-optic device functions, which could enable flexible and reconfigurable electro-optic applications in future photonic system architectures.

**Supporting Information**.

Low electric field transport at room temperature; Tailoring of thermal radiation intensity of graphene light emitter with hBN layers; Electro-thermal calculation of graphene light emitter under high electric field; Raman spectroscopy; Optical spectroscopy; Time-resolved light emission; Electrical RC time constant; Current saturation and negative differential conductance under high electric field; Transient temperature in graphene heterostructures; Thermal radiation efficiency. Supporting information is available free of charge at http://pubs.acs.org.

**Note**

The authors declare no competing financial interest.

**Acknowledgements**

Y.D.K was partially supported by the Columbia University SEAS Translational Fellow program. At Columbia, device fabrication and electrical testing were supported by the Office of Naval Research N00014-13-1-0662; optical spectroscopy was supported by DOE-BES DE-FG02-00ER45799. Ultrafast measurements at MIT were supported in part by the Center for Excitonics, an Energy Frontier Research Center funded by the U.S. Department of Energy, Office of Basic Energy Sciences under award no.DE- SC0001088. M.H.B was supported grants from the National Research Foundation of Korea (NRF-2015R1A2A1A10056103, SRC2016R1A5A1008184) funded by the Korea government. D.S. and H.C. was supported by NRF grant funded by the Korea government (No. 2017R1A2B3011586) and the third Stage of Brain Korea 21 Plus Project. A.N. and T.L was supported by a DARPA grant award FA8650-16-2-7640. K.W. and T.T acknowledge support from the Elemental Strategy Initiative conducted by the MEXT, Japan and JSPS KAKENHI Grant Number JP15K21722. The Stanford authors acknowledge support by the National Science Foundation (grant no. DMR-1411107 for Raman measurements) and by the Air Force Office of Scientific Research (grant no. FA9550-12-1-0119 for emission measurements).

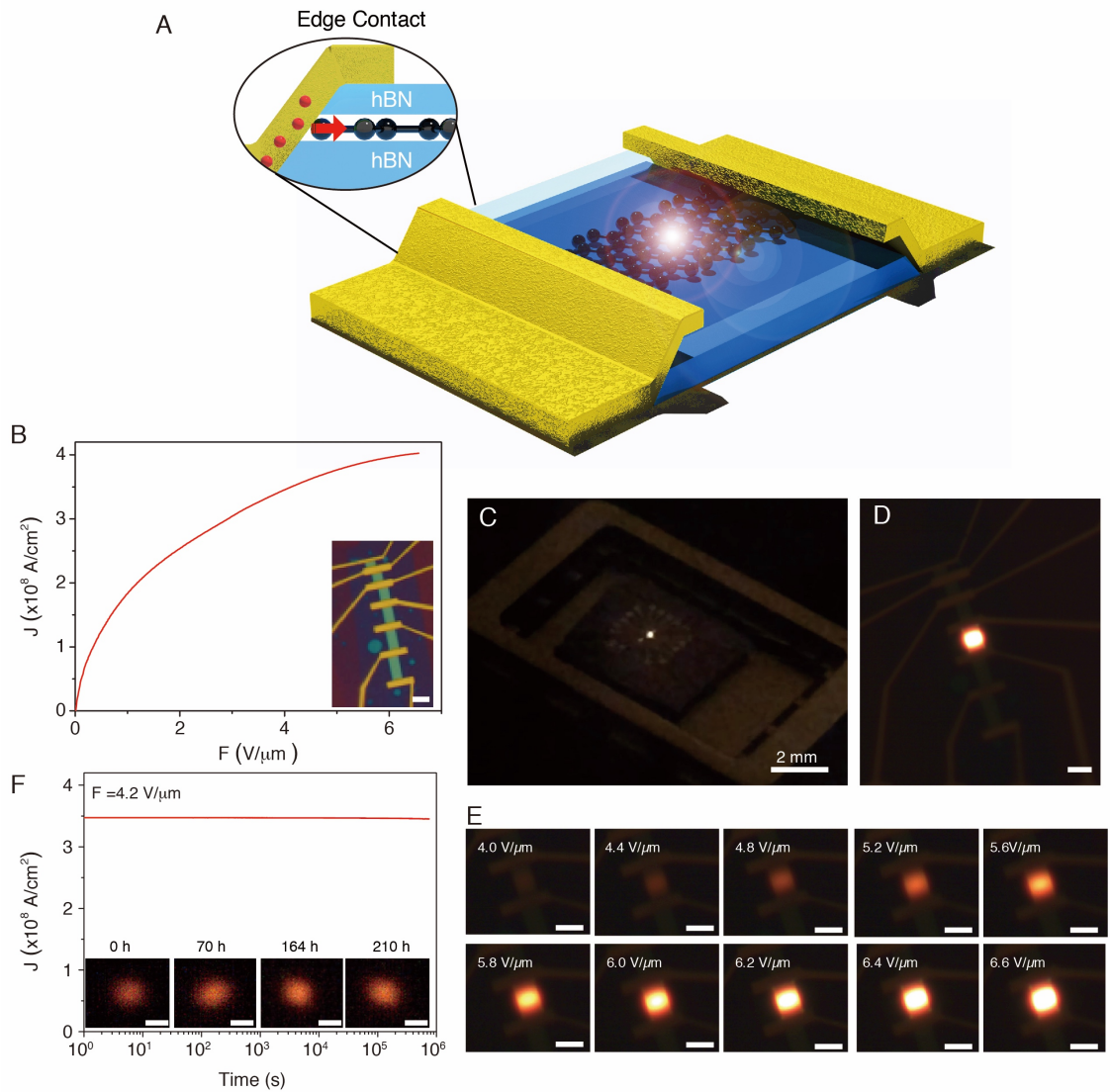

**Fig. 1. Ultrafast hBN encapsulated graphene thermal light emitter.** (**A**) The device consists of a monolayer graphene encapsulated on top and bottom by hBN; it has a one-dimensional edge contact to the drain and source contact (see inset). (**B**) Current density ($J$) as a function of applied electric field ($F$) for an emitter with channel length of 5 μm and width of 3 μm (see inset with scale bar of 6 μm). (**C**) Optical images show bright visible light emission from a microscale (3 μm - 8 μm) individual graphene light emitter under applied electric field ($F$ = 6 V/μm). (**D to E**) The graphene surface uniformly emits across the entire graphene/hBN heterostructures (3 μm – 6 μm) (D) and radiation intensity increase by the applied electric field (E). (Scale bar of 6 μm). (**F**) Long-term stability of graphene light emitter under vacuum. The current density ($J$) of graphene light emitter under constant electric field ($F$ = 4.2 V/μm) was measured during $10^6$ seconds (over 270 hours), showing negligible variation in current density and light emission intensity. Inset shows the optical images. (Scale bar of 6 μm).

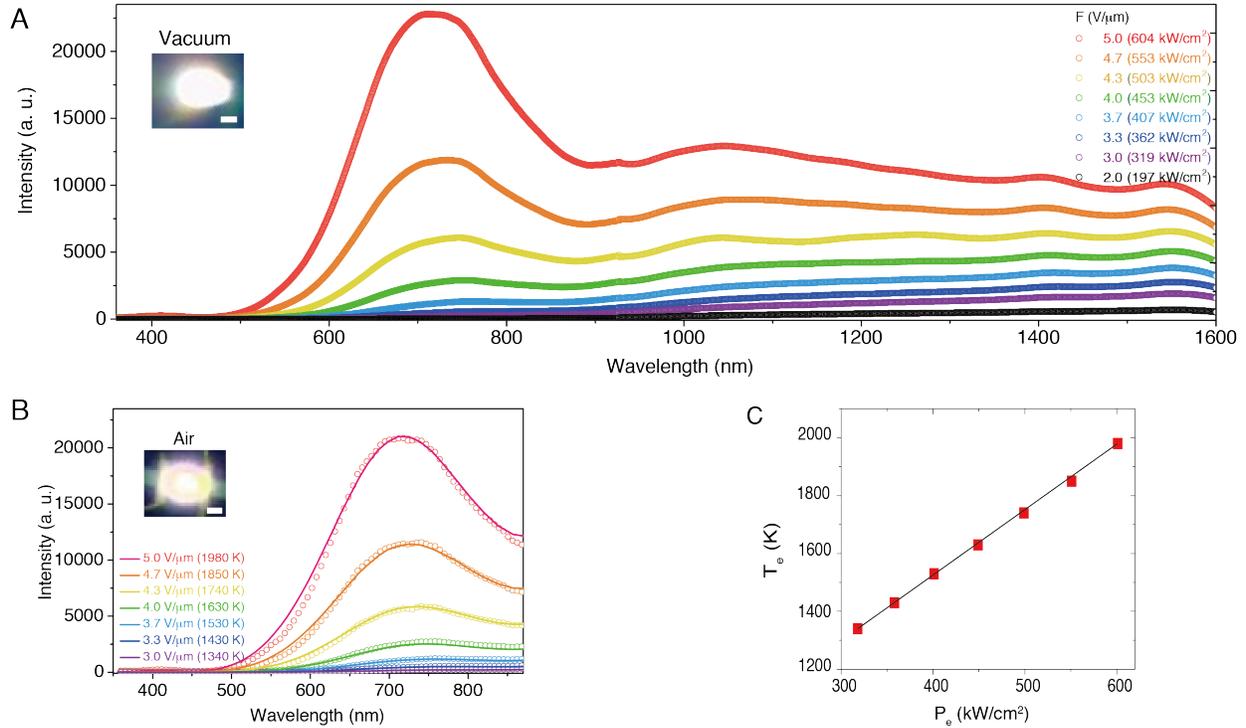

**Fig. 2. Radiation spectrum of graphene light emitter under vacuum and air. (A)** Measured radiation spectrum of graphene light emitter (scatter) under vacuum with various $F$ and electric power. We find an emission peak at the around 718 nm and a flat response at the near infrared range for high $F$ values. Inset shows the optical image of visible light emission under $F$ = 5.0 V/μm. (Scale bar is 6 μm). **(B)** Measured radiation spectrum of graphene light emitter (scatter) under air and calculated thermal radiation (solid line) based on the estimated electron temperature and grey-body thermal radiation by Plank's law with strong light-matter interaction. Inset shows the optical image of visible light emission under $F$ = 4.3 V/μm. (Scale bar is 6 μm). **(C)** $T_e$ as a function of $P_e$ under air (red square). Solid line is a linear fit to the data.

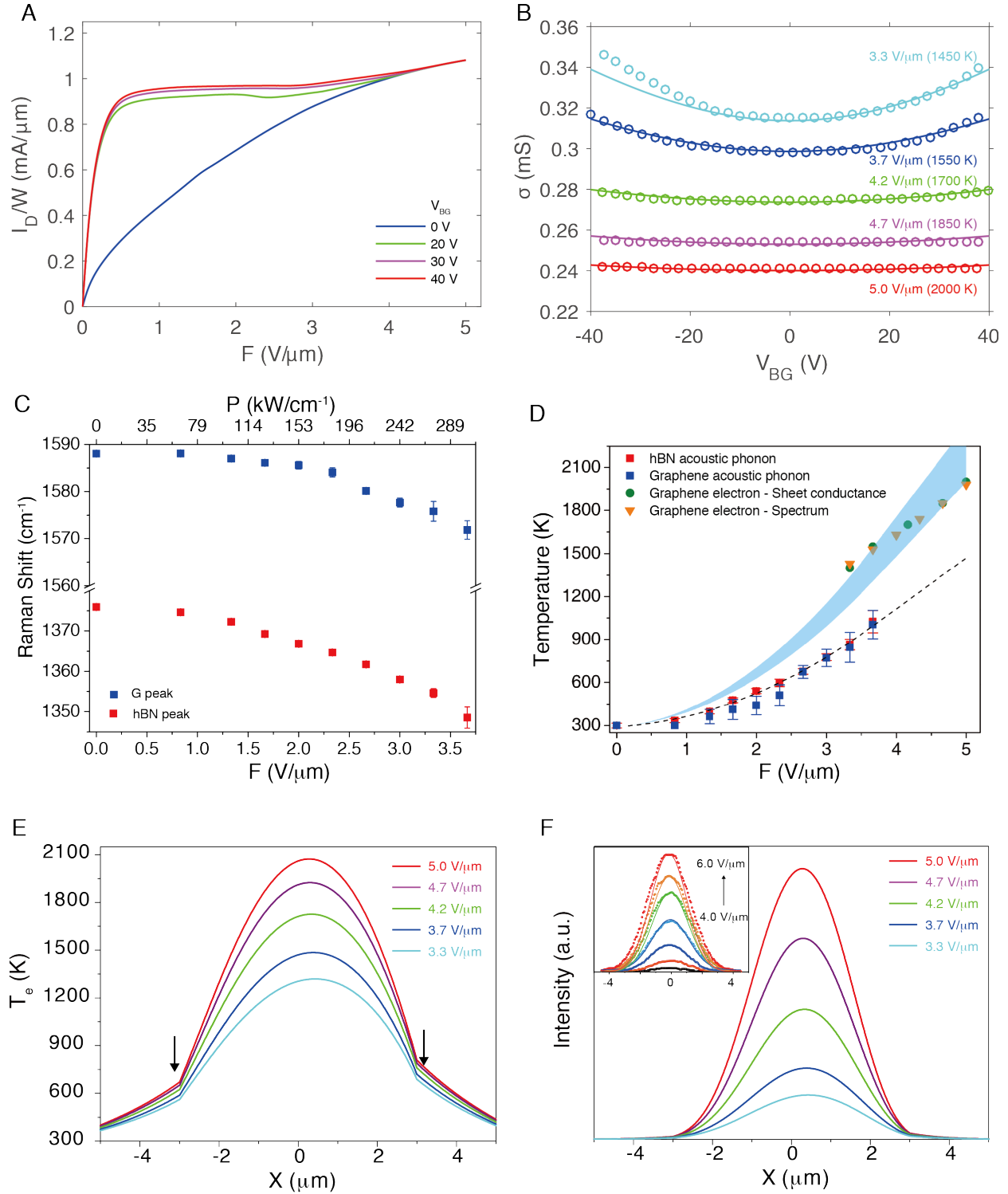

**Fig. 3. Electronic and lattice temperatures in the graphene light emitter.** (**A**) Current as function of applied electric field ($F$) for various gate voltage ($V_{BG}$). Above the critical electric field ($F > 4$ V/μm), current levels are not changed by $V_{BG}$. (**B**) Sheet conductance ($\sigma$) modulation by $V_{BG}$ of graphene heterostructure for various $F$. The electron temperature ($T_e$) is estimated based on a simulation of thermally generated charge carriers by $F$. Experimental data

(scatter) and simulation (solid line) of σ agree well. **(C)** Raman spectroscopy of graphene/hBN heterostructure to estimate the lattice temperature ($T_{ap}$). Raman peak shift of the hBN $E_{2g}$ and graphene G modes as a function of $F$. **(D)** Decoupling of electron and lattice temperature in graphene light emitters. Values of $T_e$ are calculated from the emission spectrum (orange triangles) and σ modulation (green circles), and $T_{ap}$ of graphene (blue squares) and hBN (red squares) are estimated from the Raman peak shift. The black dashed line is fitting of $T_{ap}$ and the shaded region is obtained for non-equilibrium temperature coefficients $\alpha \sim$ 0.45-0.77. **(E to F)** Calculated $T_e$ profile of the graphene light emitter for various values of $F$ (arrows indicate the edge of metal electrodes). (F) Calculated radiation intensity profile for various values of $F$ based on the temperature profile and the Stefan-Boltzmann law. (Inset, measured optical intensity profile (scatter) and Gaussian fitting (solid line) based on the optical images of Fig. 1E.)

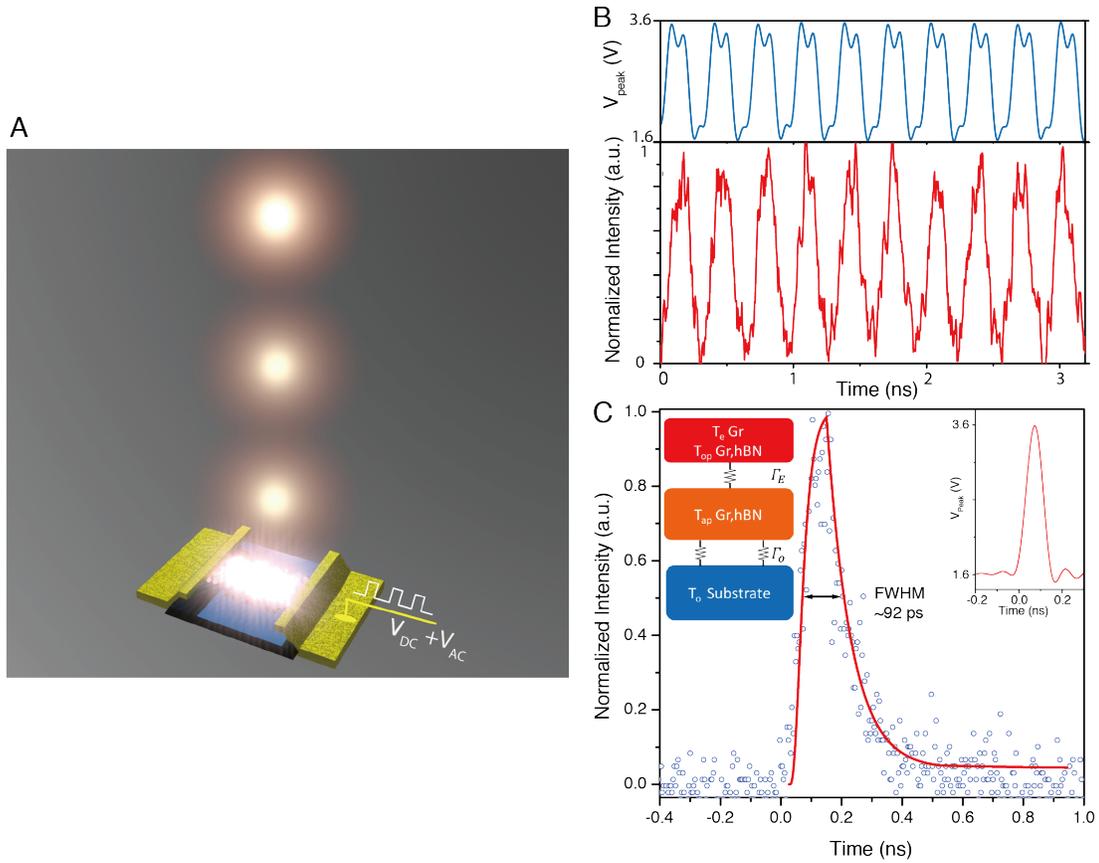

**Fig. 4. Generation of ultrafast light pulses by the electric control. (A)** Schematic of the electrically driven ultrafast graphene light emitter. The temporal profile of the light pulses are recorded by time correlated single-photon counting. **(B)** Emission profiles (lower panel) for pulsed electrical excitation (upper panel). The emission profile follws the electrical drive at the indicated frequency of ~ 3GHz. **(C)** Generation of ultrafast (92 ps) light pulses from the graphene light emitter (blue solid line) for an 80 ps electrical drive pulse, corresponding to a bandwidth of 10 GHz. According to the transit temperature and thermal radiation exponential fit (red solid lines). Insets, (Left) schematic of energy relaxation of graphene. The red block corresponds to quasi-equilibrium of electrons of in graphene and the strongly coupled optical phonons of the graphene/hBN by hybrid polaritonic modes under electrical excitation. Subsequently, the heat flows to the acoustic phonons and the substrate. (Right) Temporal profile of the 80 ps electrical drive pulse.